# Navigating the Ocean with DRL:
# Path following for marine vessels

Joel Jose[1], Md Shadab Alam[2], and Abhilash Sharma Somayajula[3]

**ABSTRACT**

Human error is a substantial factor in marine accidents, accounting for 85% of all reported incidents. By reducing the need for human intervention in vessel navigation, AI-based methods can potentially reduce the risk of accidents. AI techniques, such as Deep Reinforcement Learning (DRL), have the potential to improve vessel navigation in challenging conditions, such as in restricted waterways and in the presence of obstacles. This is because DRL algorithms can optimize multiple objectives, such as path following and collision avoidance, while being more efficient to implement compared to traditional methods. In this study, a DRL agent is trained using the Deep Deterministic Policy Gradient (DDPG) algorithm for path following and waypoint tracking. Furthermore, the trained agent is evaluated against a traditional PD controller with an Integral Line of Sight (ILOS) guidance system for the same. This study uses the Kriso Container Ship (KCS) as a test case for evaluating the performance of different controllers. The ship's dynamics are modeled using the maneuvering Modelling Group (MMG) model. This mathematical simulation is used to train a DRL-based controller and to tune the gains of a traditional PD controller. The simulation environment is also used to assess the controller's effectiveness in the presence of wind.

## 1. INTRODUCTION

Internationally reported statistics indicate that the total percentage of accidents attributable to human error is close to 85%, caused fully or in part by improper or delayed human response. Such incidents not only incur human and economic losses, but also prove to be catastrophic to the environment. The recent incident at the Suez Canal (2021) involving the Ever-Given container ship is just one among many examples where the vessel could not maintain its path under the influence of strong winds. Automation using artificial intelligence can potentially reduce the risk of accidents by eliminating human intervention in vessel navigation. Advancements in the field of AI allow for more complex methods that can better adjust to uncertainties in the environment. Recent efforts of IMO include promoting the development of a regulatory framework for Marine Autonomous Surface Ships (MASS), allowing for designers and owners to consider autonomous ships in the maritime industry.

While automated path following of marine vessels is widely explored in the light of traditional control, contemporary studies have been made into the use of reinforcement learning (RL) based methods for path following and collision avoidance. (Wang et al. 2019) used Q-learning for path following of an Autonomous ship. (Sivaraj et al. 2022) uses a DQN Agent in the control of a KVLCC2 model for path following in calm water. (Woo et al. 2019) demonstrated successful control of a USV in path following using a Deep Deterministic Policy Gradient Algorithm. (Guan et al. 2022) uses a PPO agent to perform path following and behavior decision-making of an intelligent smart marine autonomous surface ship. DRL techniques have also been used to successfully incorporate obstacle avoidance capabilities, in compliance with COLREGs. (Shen et al. 2019) successfully trains a DQN agent for automatic collision avoidance of multiple ships. (Meyer et al. 2020) demonstrated the performance of a PPO algorithm in following a given path while avoiding multiple static obstacles.

[1] IIT Madras; 0000-0002-8358-5360; na18b111@smail.iitm.ac.in
[2] IIT Madras; 0000-0001-9184-9963; oe21s007@smail.iitm.ac.in
[3] IIT Madras; 0000-0002-5654-4627; abhilash@iitm.ac.in

Although extensive research has been made on path following and obstacle avoidance capabilities of marine surface vessels, limited studies focus on the comparison between the effectiveness of these controllers with conventional methods for path following used in the industry. Therefore, the aim of this study is to develop a Deep Deterministic Policy Gradient (DDPG) agent for path following, and to access its performance with a conventional autopilot system consisting of an Integral Line-of-Sight (LOS) guidance and Proportional-Derivative (PD) controller.

The remainder of this study is organized as follows. Section 2 explains the dynamics of the numerical model used for simulation in this study. Section 3 briefly introduces the basics of deep reinforcement learning and the fundamentals of a DDPG algorithm. Section 4 explains the framework used for training the DDPG agent for path following. Section 5 assesses the performance of the trained DDPG agent in both calm waters and in presence of environmental forces. The results are compared with a conventional autopilot system subject to identical conditions. Finally, Section 6 summarizes the results and discussion of this study.

## 2. SHIP NUMERICAL MODEL

The KCS (Kriso container ship) design was conceived to provide data for validating CFD analysis for a modern container ship and was hence chosen to run numerical simulations and test the autopilot systems used in this study. The model data is readily available for research use, of which the main particulars are given in Table I.

TABLE I: MAIN PARTICULARS

| Main Particulars | Full Scale | 1:75.5 Scale |
|---|---|---|
| **Hull** | | |
| Lpp (m) | 230.0 | 3.0464 |
| D (m) | 19.0 | 0.2517 |
| T (m) | 10.8 | 0.143 |
| Displacement ($m^3$) | 52030 | 0.1209 |
| **Propeller** | | |
| Type | Fixed pitch | Fixed pitch |
| No. of blades | 5 | 5 |
| Diameter (m) | 7.9 | 0.105 |
| **Rudder** | | |
| Type | Semi-balanced horn rudder | Semi-balanced horn rudder |
| Profile | NACA 0018 | NACA 0018 |
| Aspect Ratio | 1.8 | 1.8 |
| **Test Conditions** | | |
| LCG($x_G$) | 111.6 | 1.478 |
| Service Speed (U) (m/s) | 24 | 1.1 |

The ship dynamics are mathematically modelled using the MMG (Maneuvering Modelling Group) model proposed in (Yoshimura and Masumoto 2012, Yasukawa and Yoshimura 2015). Three non-linear equations of motion are used to solve for the vessel's motion in surge, sway, and yaw directions. Based on a given rudder command $\delta_c$, the equations of motion are solved progressively at each time step as an initial value problem using a Runga-Kutta explicit solver.

Two coordinate frames are defined to model the dynamics of the vessel: The global coordinate system (GCS) and the body-fixed coordinate system (BCS). The GCS is an earth fixed coordinate frame with the z-axis pointed down, used to visualize the ship's trajectory. The body-fixed reference frame is fixed to the current position of the vessel in the global frame. The axes are aligned such that the x-axis points to the vessel's bow, y-axis to the starboard side of the vessel, and z-axis pointing down. The origin of the frame coincides with the point of intersection of the midship, centerline and waterline of the vessel. The above two frames can be visualized in Fig. 1

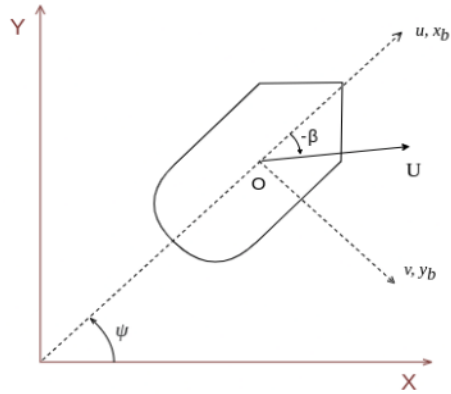

Figure.1: GCS and BCS frames

The position and orientation of the vessel in the global frame is denoted by $\eta = [x_o, y_o, \psi]^T$ where $x_o, y_o$ is the position of the origin of BCS expressed in GCS, and $\psi$ is the heading angle, defined as the angle about the global z-axis between the positive x-axes of the GCS and BCS. Given the heading angle, the rotation matrix from BCS to GCS is given by:

$$[R(\psi)] = \begin{bmatrix} cos(\psi) & -sin(\psi) & 0 \\ sin(\psi) & cos(\psi) & 0 \\ 0 & 0 & 1 \end{bmatrix} \qquad (1)$$

The velocity vector of the vessel is denoted as $V = [u, v, r]^T$, where $u$, $v$ and $r$ are surge, sway, and yaw velocities of the vessel respectively. The total speed, $U$ of the vessel is given by $U = \sqrt{u^2 + v^2}$. The drift angle, $\beta$ is defined as the angle measured from the x-axis of the BCS to the total velocity $U$, measured about global z-axis, computed as $\beta = tan^{-1}(-v/u)$. The vessel kinematics can therefore be expressed as below:

$$\dot{\eta} = [R(\psi)]V \qquad (2)$$

In order to generalize simulation results to any given scale of the vessel, the MMG model equations are non-dimensionalized in accordance with prime-II system of normalization as described by (Lekkas and Fossen 2012). For this purpose, surge and sway equations are non-dimensionalized by dividing both sides of the surge and sway equations by $(\rho U^2 LT)/2$, where $\rho$ is the density of sea water, $L$ is the length of the vessel, $U$ is the design speed of the vessel, and *T* is the design draft of the vessel. Similarly, the yaw equation of motion is non-dimensionalized by dividing both sides by $(\rho U^2 LT)/2$. The non-dimensional mass and added mass terms are specified in Table 2.

TABLE 2: SHIP PARAMETERS

| Parameter | Non-dimensional value |
|---|---|
| Surge added mass $(m_x)$ | 0.006269 |
| Sway added mass $(m_y)$ | 0.155164 |
| Yaw added mass moment $(J_{zz})$ | 0.009268 |
| Yaw mass moment of inertia $(I_{zz})$ | 0.011432 |
| Mass of the vessel (m) | 0.18228 |

In order to simulate a more realistic response of the ship's rudder, the rudder variation rate $\dot{\delta}$ is constrained to a first order system, capped to an upper limit of $\dot{\delta}_{max}$ (5°/sec) as suggested by (Deraj et al. 2023).

## 3. DEEP REINFORCEMENT LEARNING

Reinforcement Learning (Sutton and Barto 2018) is a branch of machine learning which involves training an agent to choose optimal actions when interacting with an environment. This optimal behavior is learned through a balance between exploration and exploitation, with the agent accumulating knowledge by exploring the environment and exploiting based on the information it has learned. A typical iteration of training involves the agent performing an action $a_t$ at time *t*, based on the current state $s_t$ and policy $\pi$. Depending on the state transition, a reward $r(s_t, a_t)$ is generated by the environment, which is utilized to update the policy. RL environments are commonly defined in the form of one-step Markov Decision Processes (MDP), where the current state of the depends only on the state and the action chosen at the preceding time step.

Classical, tabular solution methods used in RL, such as Q-learning, often fall short in efficiency when applied to continuous space and control tasks. The use of function approximators, especially deep neural networks, as policies has become the norm in solving high dimensional and complex environments.

### 3.1 DEEP DETERMINISTIC POLICY GRADIENT (DDPG)

Deep Deterministic Policy Gradient (DDPG) (Lillicrap et al. 2015) is a DRL algorithm that uses the actor-critic framework to extend Q-learning to a continuous action space. The policy and Q-value functions are both estimated by neural networks, namely the actor and critic networks respectively. In policy gradient methods, the actor network directly represents the agent's policy. The actor network takes as input the current state and outputs the action based on the policy encoded by the network. The critic network takes input both the state and action to predict the Q-value. DDPG also makes use of target networks like the DQN algorithm.

In DDPG, noise is added to the action to aid the agent to explore more during training. The behavioral policy can be denoted as:

$$a_t = \mu(s_t) + N_t \qquad (3)$$

Where $N_t$ is the noise added to the policy and $\mu(s_t)$ is the current policy. Noise is usually added through a correlated Ornstein-Uhlenbeck process or an uncorrelated Gaussian distribution.

## 4. IMPLEMENTATION OF DDPG ALGORITHM FOR SHIP NAVIGATION

### 4.1 Observation State Space

At each time step, the agent receives a vector of variables known as the observation state, which reflects the agent's present state. The agent utilizes this information to determine which action to take. The states are:

$$s_t = [d_c, \chi_e, d_{wp}, r] \tag{4}$$

where, $d_c$ is the cross-track error, $\chi_e$ denotes the course angle error, $d_{wp}$ is the distance to destination and $r$ is the yaw rate of the vessel.

### 4.2 Action space

The action available for the agent to choose at any given time step is the commanded rudder angle, denoted as $\delta_c \in [-35°, 35°]$.

### 4.3 Reward Structure

In RL, a reward is a scalar value that the agent receives after taking an action in the environment. Rewards help agents to learn an optimal action at each time step. The agent's objective is to maximize the cumulative sum of rewards over time, also known as the return.

$$\begin{aligned} r_1 &= 2\, exp\left(\frac{-d_c^2}{12.5}\right) - 1 \\ r_2 &= 1.3\, exp\left(-10|\chi_e|\right) - 0.3 \\ r_3 &= \frac{-d_{wp}}{4} \end{aligned} \tag{5}$$

Eq. (5) shows the rewards associated with the training of RL agent. The rewards are associated with cross track error, course angle error and yaw rate.

$$\begin{aligned} r_t &= r_1 + r_2 + r_3 \\ R &= \Sigma_0^t r_t \end{aligned} \tag{6}$$

In eq. (6), $r_t$ is the reward at time step $t$ and $R$ is the episode return, which is the cumulative sum of rewards obtained at each time step.

### 4.4 Training Process

At the start of each training episode, the ship has an initial velocity in the surge direction with no velocity in sway and yaw direction. The destination point is chosen randomly between 8 to 28 ship lengths from the vessel's initial position at a random angle. The following termination conditions are evaluated at each timestep for ending an episode:

- *Successful termination:* A training episode is considered successful if the vessel enters a tolerance region of 0.5L surrounding the destination waypoint. A positive terminal reward of +100 is granted in this case.

- *Time limit condition:* The agent has a maximum of 160 time-steps to reach the destination. If it fails to do so within this time, the episode ends unsuccessfully.

- *Missed waypoint condition:* A missed waypoint condition is triggered when the ship passes beyond the destination waypoint, while its velocity vector points away from it (indicating that the vessel is moving away from the destination). In this case, the episode ends unsuccessfully.

### 4.5 Hyperparameters of the networks

The agent is tuned by experimenting with different configurations for the actor, critic networks and other hyperparameters. The actor network as well as critic network both use 2 hidden layers with *tanh* activation in the hidden layers. After each time step, the transition is stored in the replay buffer for updating the neural network. The maximum replay buffer size is set to be 100000 and it follows FIFO (First In First Out) method.

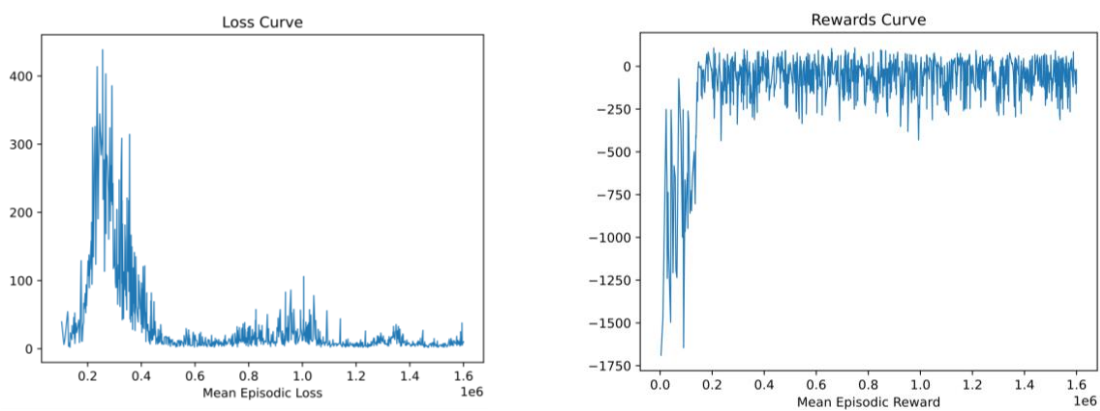

Fig.2: Returns and Loss plot for DDPG agent.

The best set of hyperparameters is given in Table I. Fig. 2 shows the plot of the returns and loss in every iteration during training.

TABLE I: Hyperparameters for DDPG Model

| Hyperparameter | Value |
|---|---|
| Learning Rate | 0.0008 |
| Activation Function | tanh |
| Optimizer | Adam |
| Actor Hidden Layers | 64,64 |
| Critic Hidden Layers | 64,64 |
| Discount Factors | 0.97 |
| Batch Size | 32 |
| Noise Multiplier | $\sigma = 0.0698; \mu = 0$ |
| Update Frequency | 10 |
| Trained Duration | 1280000 |
| Soft Update Factor | 0.02 |

### 5. RESULTS

After training the agent for sufficient time steps, the trained agent is evaluated through various waypoint tracking scenarios. In this section, the agent robustness is first checked through waypoint tracking in different quadrants, followed by giving complex paths that are discretized into number of waypoints. Finally external disturbances such as winds are introduced in the environment and the agent is asked to follow these complex paths.

## 5.1 Waypoint Tracking

To evaluate the DDPG agent's performance, destination waypoints were assigned in each of the four quadrants: (10L, 10L), (-10L, 10L), (10L, -10L), and (-10L, -10L). The vessel's starting point was set at the origin with a surge velocity of 1 and a heading angle ($\psi$) of 0. Fig. 3 shows that the trained model reliably tracks the destination waypoint in all quadrants.

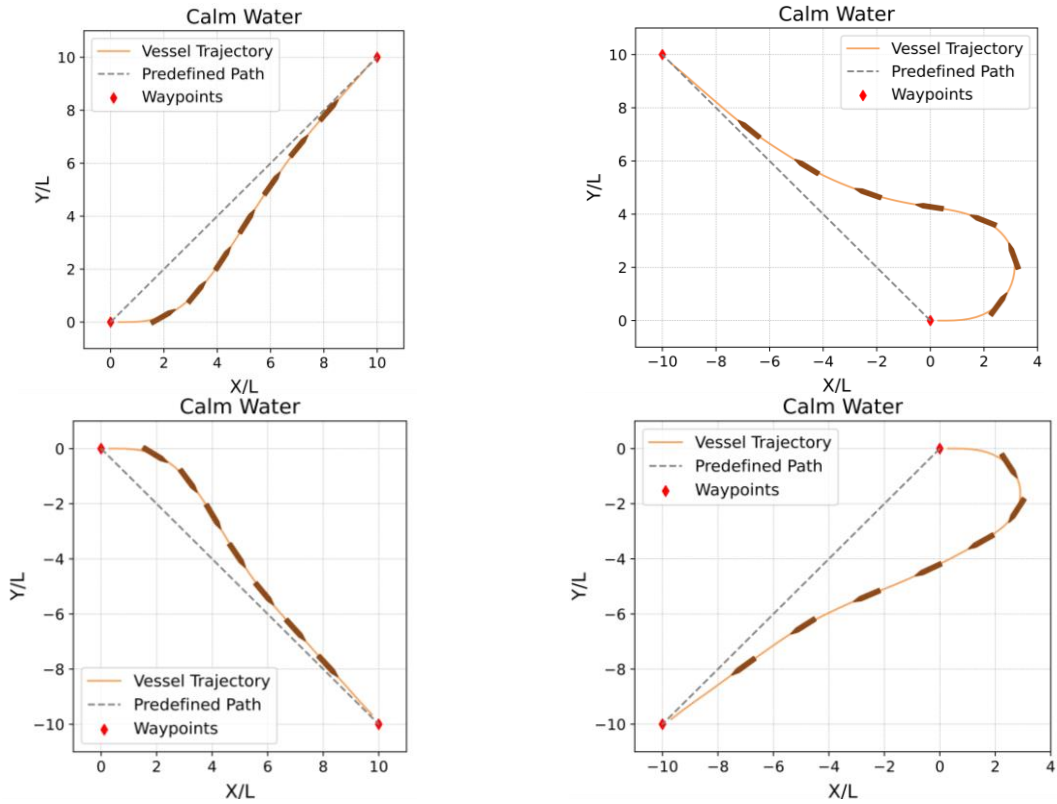

Fig.3: Single waypoint tracking

As seen in Fig. 3, the model was successfully able to track waypoints in each quadrant.

## 5.2 Path following through waypoint tracking.

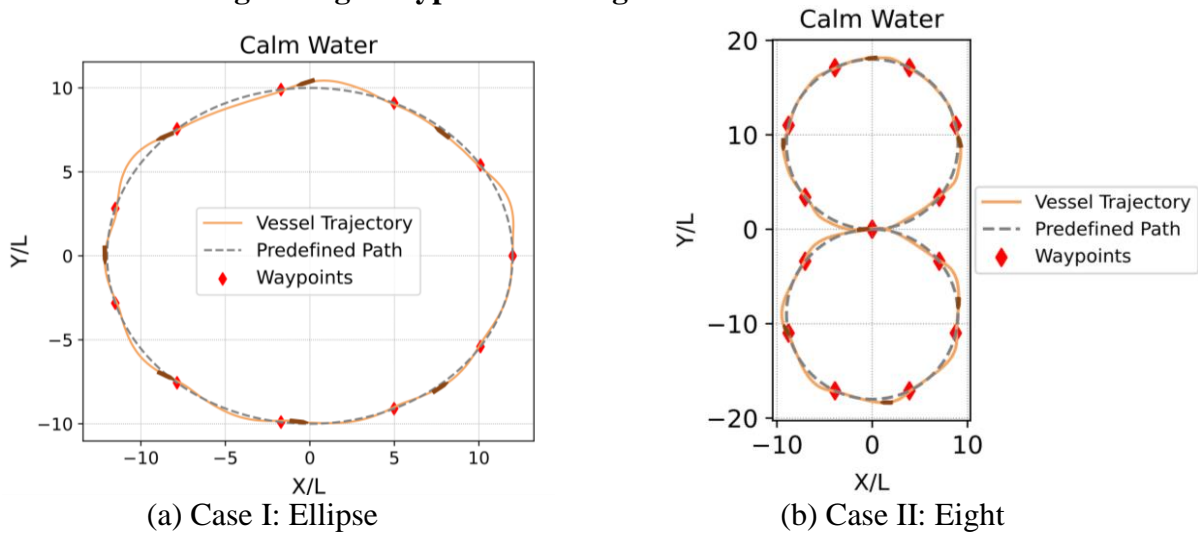

(a) Case I: Ellipse          (b) Case II: Eight

Fig.4: Path Following cases.

The following trajectories are defined for further evaluation which has been discretized into several waypoints. Fig. 4(a) shows an elliptical trajectory which is discretized into 15 waypoints. The length of the semi-major axis and semi minor axis of the ellipse is *14L* and *10L*. The ship starts from point (14L, 0) with a heading angle of $\pi/2$ and a surge velocity of *1U*. Similarly, as shown in Fig. 4(b) a shape of numeral '8' is chosen and discretized into 23 waypoints. In this case, the ship starts at point (0, 0) and the heading angle is 0 with a surge velocity of *1U*. The radius of the circle in the trajectory is *9L*.

## 5.3 Performance with wind forces

Wind forces and moments are modelled and included as suggested by (Deraj et al. 2023). Fig. 4 shows two specific cases that were examined, each with different wind speeds and directions. Even though the DDPG agent was trained only on calm water conditions, it can be inferred from the figures that the trained agent is able to compensate for external disturbances and successfully track the waypoints.

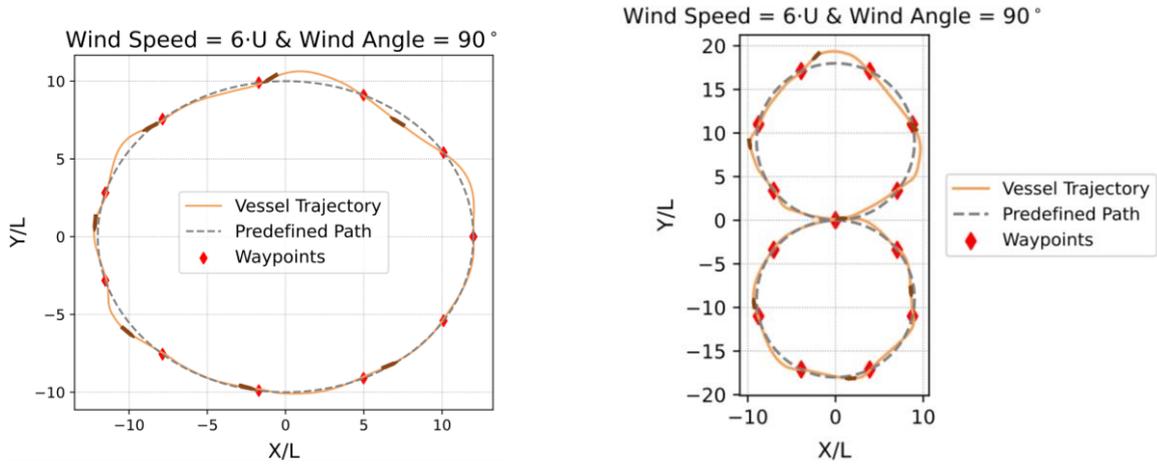

Fig.5: Path following cases in presence of constant and uniform wind.

## 6. DISCUSSION

### 6.1 Comparison with a conventional autopilot system

In this section, a comparison is made between the DDPG agent's ability to follow a path with that of a conventional autopilot system consisting of a PD controller and ILOS guidance system. A PD controller ensures that the vessel's heading angle $\psi$ converges to a desired heading angle $\psi_d$. The desired heading angle $\psi_d$ is obtained using the integral line of sight (ILOS) guidance law as described by (Fossen 2021). The error $e$ is determined as the difference between the current heading angle $\psi$ and the reference value $\psi_d$, and is expressed in eq. 7(a). Similarly, eq. 7(b) represents the time derivative of the error.

$$e = \psi_d - \psi \quad (7(a))$$

$$\dot{e} = \dot{\psi}_d - \dot{\psi} = -\text{r} \quad (7(b))$$

Thus, the proportional derivative (PD) control law can be expressed as

$$\delta_c = K_p e + K_d \dot{e} = K_p(\psi_d - \psi) - K_d r \quad (8)$$

where $\delta_c$ represents the commanded rudder angle

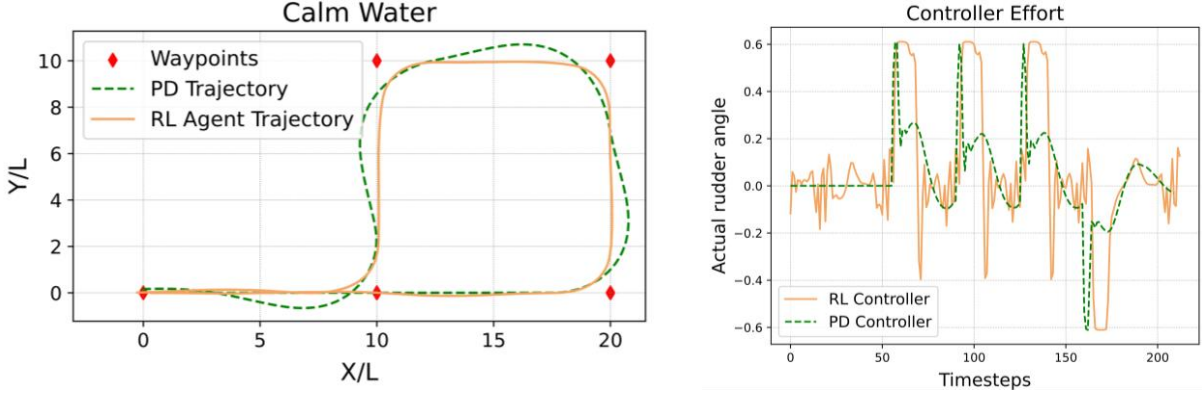

Fig. 6: Comparison between DDPG agent and a PD controller in calm water.

$K_p$ and $K_d$ corresponds to the proportional and derivative gains of the PD controller respectively. These gains are adjusted to minimize the deviation of the vessel's trajectory from the desired trajectory. In addition, an ILOS guidance algorithm utilizes a look-ahead distance parameter Δ and an integral gain $k$, which determines the vessel's path tracking behavior.

After tuning the controller and guidance laws, $K_d$ = 4.0 and $K_p$ = 2.5 for the PD controller gains, Δ = 2L and $k$ = 0.05 for the lookahead distance and integral gain of the ILOS guidance respectively were found to provide the best performance for the chosen vessel.

Fig. 6 shows that the DDPG controller performs slightly better than the PD controller in terms of the RMS of cross track error with an improvement of 8% compared to the PD controller. However, the DDPG agent imposes a higher controller effort than the PD controller as well.

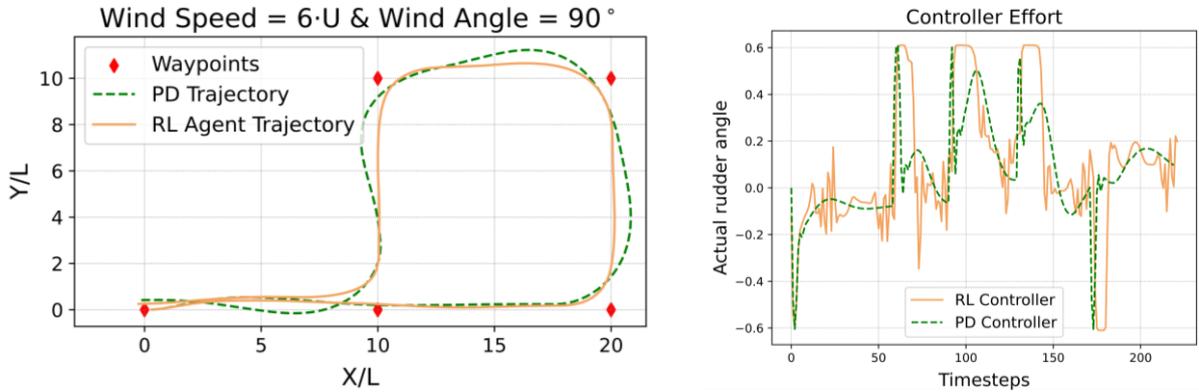

Fig. 7: Comparison between DDPG agent and a PD controller in presence of wind.

Fig. 7 compares the performance between the DDPG agent and the conventional system in the presence of wind. A similar improvement of 7.5% in the RMS of cross track error can be observed with the DDPG agent here as well, but at the cost of more controller effort.

## 7. CONCLUSION

The findings of this study demonstrate the effectiveness of utilizing a DRL-based controller, specifically a DDPG agent, for path following of a ship through waypoints. By training the DDPG agent, the agent was able to successfully navigate through complex paths that were represented by waypoints. The trained DDPG agent seems to demonstrate slightly better path following behavior compared to a conventional autopilot system. However, this comes at a cost on the rudder controller effort which can be minimized with better tuning of hyperparameters.

Further research in this area may involve exploring the use of more advanced DRL algorithms and improved reward structures that could improve the performance of the controller, while minimizing controller effort. Furthermore, the current DRL framework will be enhanced to include obstacle and collision avoidance, as well as adherence to COLREGs, and a subsequent comparison with traditional collision avoidance methods employed in the marine industry.

## 8. ACKNOWLEDGEMENT

This work was partially funded by the Science and Engineering Research Board (SERB) India - SERB Grant CRG/2020/003093 and New Faculty Seed Grant of IIT Madras. This work is also supported through the proposed {Center of Excellence for Marine Autonomous Systems (CMAS), IIT Madras} setup under the Institute of Eminence Scheme of Government of India.